\documentclass[aps,showpacs,prd]{revtex4}
\usepackage{graphicx}
\usepackage[all]{xy}
\usepackage{amsmath}
\usepackage{amssymb}
\usepackage{accents}
\usepackage[usenames,dvipsnames]{color}
\usepackage[latin1]{inputenc}

\newcommand{\be}{\begin{equation}}
\newcommand{\ee}{\end{equation}}
\newcommand{\ben}{\begin{eqnarray}}
\newcommand{\een}{\end{eqnarray}}
\newcommand{\bes}{\begin{subequations}}
\newcommand{\ees}{\end{subequations}}

\newcommand{\bb}{\bibitem}
\newcommand{\bn}{\begin{eqnarray}}
\newcommand{\en}{\end{eqnarray}}

\begin{document}

\title{A general method for transforming non-physical configurations in BPS states}
\author{J. R. L. Santos$^{a}$, A. de Souza Dutra$^{b}$, O. C. Winter$^{b}$, R. A. C. Correa$^{c,d}$}

\affiliation{$^a$Unidade Acad\^{e}mica de F\'{\i}sica, Universidade de Federal de Campina Grande, 58109-970 Campina Grande, PB, Brazil}

\affiliation{$^b$S\~ao Paulo State University (UNESP), Campus de Guaratinguet\'a-DFQ, Avenida Dr. Ariberto Pereira da Cunha 333, 12516-410 Guaratinguet\'a, SP, Brazil}

\affiliation{$^c$Instituto Tecnol\'{o}gico de Aeron\'{a}utica, DCTA, 12228-900, S\~{a}o Jos\'{e} dos Campos, SP, Brazil}
\affiliation{$^d$Scuola Internazionale Superiore di Studi Avanzati (SISSA)
via Bonomea, 265, I-34136 Trieste, Italy}

\begin{abstract}
In this work, we apply the so-called BPS method in order to obtain topological defects for a complex scalar field Lagrangian introduced by Trullinger and Subbaswamy. The BPS approach led us to compute new analytical solutions for this model.{\bf  In our investigation, we found analytical configurations which satisfy the BPS first-order differential equations but do not obey the equations of motion of the model. Such defects were named non-physical ones. In order to recover the physical meaning of these defects, we proposed a procedure which can transform them into BPS states of new scalar field models. }The new models here founded were applied in the context of hybrid cosmological scenarios, where we derived cosmological parameters compatible with the observed Universe. Such a methodology opens a new window to connect different two scalar fields systems and can be implemented in several distinct applications such as Bloch Branes, Lorentz and Symmetry Breaking Scenarios, Q-Balls, Oscillons, Cosmological Contexts, and Condensed Matter Systems.
\end{abstract}

\pacs{11.10.Lm, 03.50.-z, 02.30.Jr, 05.45.Yv}

\maketitle

\section{Introduction}

Topological defects are present in several scenarios of physics, covering areas like braneworld models, quintessence cosmological approaches, condensed matter, among others \cite{vachaspati, koley, slatyer, bg, kinney, bglr, bllm, dsw}. As examples of the applicability of defects solutions, we refer to studies involving defects in massive integrable field theories in 1+1 dimensions \cite{JHEP2017}, in systems where the Lorentz symmetry is violated \cite{PRD2011, AOP2015}, in 2D materials \cite{AdP2018}, and in Yang monopoles \cite{PRL2018}.

  A well-established method to determine defect-like solutions is the so-called BPS method, proposed by  Bogomol'ny, Prasad, and Sommerfeld \cite{bps}. Such a method is based on the assumption that the fields obey first-order differential equations (BPS differential equations), in order to minimize their energy density. Therefore, the solutions corresponding to these first-order differential equations (here also called BPS solutions) must satisfy the equations of motion of given systems. The main advantage of the BPS method is that one needs to deal with first-order equations instead of second or higher order differential equations. Such a method has been mainly applied in the context of Lagrangian densities composed by real scalar fields, by complex scalar fields, and by gauge field theories, as one can see in references \cite{alonso_02,dutra_plb,montonen,bb_00,bazeia_vortice}. Generalizations of the BPS approach can be found in the literature as in Ref. \cite{bsr95}, where the authors show how the energy of the defects can saturate to a bound energy if different sets of BPS differential equations were obeyed. In this paper, we are going to explore the application of the BPS method in a complex field model proposed by Trullinger and Subbaswamy \cite{trull}.

{\bf At the end of the seventies, Trullinger and Subbaswamy found topological solutions (or defects) which satisfy the equations of motion related with the following Lagrangian density}
\be \label{in_eq1}
{\cal L}=\frac{1}{2}\left|\psi_t(x,t)\right|^2-\frac{1}{2}\left|\psi_x(x,t)\right|^2+\frac{A}{2}\left|\psi(x,t)\right|^2
-\frac{B}4|\psi(x, t)|^4-D\,\left|\psi(x,t)\right|^N\,\left\{1-\cos[N\,\phi(x,t)]\right\}\,,
\ee
with $\psi(x,t)=u(x,t)\,e^{i\,\phi(x,t)}$, where $A$, $B$, and $D$ are all positive constants. In their work, Trullinger and Subbaswamy {\bf analyzed the case  $N=4$,} since this model has some interesting physical motivations in studies involving anisotropic ferromagnets \cite{trull}, where the defects are analogous to magnetic domain walls. {\bf The mentioned domain walls were applied in the context of condensed matter physics and in statistical mechanics.}

In this paper we are going to investigate the model introduced in \cite{trull} from the point of view of the BPS approach, moreover, we will show carefully what are the conditions to have BPS solutions which satisfy the equations of motion and the first-order differential equations for such a model. Despite the success of this methodology, we are going to unveil new sets of solutions for the first-order differential equations which do not satisfy the equations of motion coming from Eq. \eqref{in_eq1}. {\bf These solutions also have energy different from the BPS sector of the Lagrangian above, and we named them as non-physical ones. Our main objective in this paper is to transform such non-physical solutions into new BPS states for other effective models, recovering their physical meaning}. This approach also shows a new type of connection between scalar fields models which have not been observed in the literature, as far as we know.

In order to show the potential of our methodology, we apply the derived models in the context of hybrid cosmological scenarios. Since the seminal work of Kinney in this subject \cite{kinney}, several approaches to deal with cosmological models driven by more than one scalar field have been proposed \cite{bglr,ms_prd,multi_18}. The desire for hybrid models increased in the last few years after the work of Ellis {\it et al.} \cite{ellis}, where the authors unveiled that models composed by several scalar fields are compatible with the scalar index and with the tensor-to-scalar ratio parameters found by PLANCK collaboration \cite{planck16,planck18}. 

The ideas behind this investigation are divided into the following sections: section \ref{sec_1} presents the main calculations of reference \cite{trull}, starting by rewriting the complex scalar field Lagrangian in terms of a two real scalar fields model and then finding its second-order equations of motion. In section \ref{sec_2} we determine the solutions obtained by Trullinger and Subbaswamy via the BPS approach, we also point what conditions the BPS solutions would have in order to satisfy the equations of motion. Furthermore, we determine the non-physical solutions related to the first-order differential equations for this model. In section \ref{sec_3} we show the methodology responsible to relate non-physical solutions with new sets of two-field models. Sections \ref{sec_3_1} and \ref{sec_3_2} are dedicated to the application, and to the interpretation of the derived models in the context of cosmology. Our final remarks and perspectives are shown in section \ref{sec_4}.

\section{Generalities}
\label{sec_1}

{\bf In this section, we briefly review some generalities proposed by Trullinger and Subbaswamy \cite{trull} for the treatment of the Lagrangian density presented in \eqref{in_eq1}}.  Let us start the procedures by rewriting the field  $\psi(x,t)$  as
\be
\psi(x,t)=\text{Re}\,[\psi]+i\,\text{Im}\,[\psi]=\widetilde{\xi}(x,t)+i\,\widetilde{\eta}(x,t)\,.
\ee

Then, by substituting the above relation into Eq. \eqref{in_eq1}, we obtain the following pair of Euler-Lagrange coupled equations

\be
\widetilde{\xi}_{tt}-\widetilde{\xi}_{xx}-A\,\widetilde{\xi}+B\,\widetilde{\xi}^3+(B+16\,D)\,\widetilde{\xi}\,\widetilde{\eta}^2=0\,,
\ee
\be
\widetilde{\eta}_{tt}-\widetilde{\eta}_{xx}-A\,\widetilde{\eta}+B\,\widetilde{\eta}^3+(B+16\,D)\,\widetilde{\eta}\,\widetilde{\xi}^2=0\,.
\ee

{\bf Let us work with the following redefinitions:} $\widetilde{\xi}(x,t)\rightarrow\widetilde{\xi}(s)$, and $\widetilde{\eta}(x,t)\rightarrow\widetilde{\eta}(s)$, where $s=\widetilde{\gamma}\,(x-v\,t)$, $\widetilde{\gamma}=(1-v^2)^{1/2}$, and $|v|<1$. 
At this point, we emphasize that the new variable $s$ is known in the literature as travelling one, which is very useful in the analytical treatment of nonlinear systems. The previous procedures yield to
\be
\widetilde{\xi}_{ss}+A\,\widetilde{\xi}-B\,\widetilde{\xi}^3-(B+16\,D)\,\widetilde{\xi}\,\widetilde{\eta}^2=0\,,
\ee
\be
\widetilde{\eta}_{ss}+A\,\widetilde{\eta}-B\,\widetilde{\eta}^3-(B+16\,D)\,\widetilde{\eta}\,\widetilde{\xi}^2=0\,. 
\ee

Now, the scalar fields and the variable $s$ can be rescaled as follows
\be
\xi=\frac{\widetilde{\xi}}{(A/B)^{1/2}}\,, \qquad \eta=\frac{\widetilde{\eta}}{(A/B)^{1/2}}\,, \qquad \rho=s\,A^{1/2}\,,
\ee
resulting in
\be \label{sordem_1}
\xi_{\rho\,\rho}+\xi-\xi^3-\lambda\,\xi\,\eta^2=0\,,
\ee
\be \label{sordem_2}
\eta_{\rho\,\rho}+\eta-\eta^3-\lambda\,\eta\,\xi^2=0\,,
\ee
where $\lambda\equiv 1+16D/B$. 

Looking at the previous results, it is natural to think that the Eqs. \eqref{sordem_1} and \eqref{sordem_2} could be derived from a two-field Lagrangian density with the form
\be
{\cal L}=-\frac{1}{2}\,(\xi_\rho^2+\eta_\rho^2)-V(\xi,\eta)\,,
\ee
where the scalar potential $V(\xi,\eta)$  can be written as
\be \label{pot_1}
V(\xi,\eta)=-\frac{1}{2}\,(\xi^2+\eta^2)+\frac{1}{4}\,(\xi^4+\eta^4)+\frac{\lambda}{2}\,\xi^2\eta^2\,.
\ee


 Note that, the potential \eqref{pot_1} has four symmetric degenerated minima  $M_{i}=(\xi_{i}, \eta_{i})$, which are localized in $M_{1}=(0,1)$, $M_{2}=(0,-1)$, $M_{3}=(1,0)$ and $M_{4}=(-1,0)$. Such minima are known as the vacua of the topological configurations, and the solutions connecting different topological configurations were named as $\pi/2$ and $\pi$ solutions \cite{trull}. The $\pi/2$ solution connects the topological sectors where
\ben
&&
\xi=0\qquad \eta=1 \qquad \text{for} \qquad \rho=-\infty \\ \nonumber
&&
\xi=1\qquad \eta=0 \qquad \text{for} \qquad \rho=+\infty  \\ \nonumber
&&
\xi_\rho(\pm\infty)=\eta_\rho(\pm\infty)=0\,, \\ \nonumber 
\een
while the $\pi$ solution is related with the vacua
\ben
&&
\xi=-1\qquad \eta=0 \qquad \text{for} \qquad \rho=-\infty \\ \nonumber
&&
\xi=1\qquad \eta=0 \qquad \text{for} \qquad \rho=+\infty  \\ \nonumber
&&
\xi_\rho(\pm\infty)=\eta_\rho(\pm\infty)=0\,. \\ \nonumber 
\een

In the literature about defects, one-dimensional solutions which connects two distinct vacua are called kinks, besides one-dimensional solutions related with only one vacuum are named lumps. In two scalar fields models, these one-dimensional defects are combined to construct an orbit in the field space. {\bf Therefore, the $\pi$ solutions have orbits formed by the combinations of a kink with a lump defect, while the orbits of  $\pi/2$ solutions are constructed by combinations of two kink-like solutions.}

In \cite{trull} the authors determined two analytical solutions for the case $\lambda=3$, by directly integrating their equations of motion. However, it is possible to use the so-called BPS method to generalize such solutions \cite{bps}. As we know, since this approach allows to obtain a first-order differential equation from the total energy, such insight becomes a powerful tool to solve nonlinear problems analytically.

\section{BPS Treatment}
\label{sec_2}

An advantage of the BPS method is that it simplifies considerably the integration process of equations of motion, and it also yields to new sets of analytical solutions which satisfy the BPS first-order differential equations. In this section, we will show that most part of the BPS solutions from \cite{trull} are not going to obey its equations of motion. Therefore, we will need to find models which are satisfied by these new sets of analytical solutions. In order to determine such models, we are going to use the methodology to construct scalar fields systems presented in \cite{bls}\,.

From now on, we will be dealing with the problem of obtaining a BPS bound for the model under investigation. The first step to implement the BPS method to this context consists in rewrite \eqref{pot_1} as
\be \label{pot}
\widetilde{V}(\xi,\eta)\equiv V(\xi,\eta)+\frac{1}{4}=
\frac{1}{2}\,\left[\frac{1}{\sqrt{2}}\,(\xi^2+\eta^2-1)\right]^2+\frac{1}{2}\,\left[\sqrt{\lambda-1}\,\xi\,\eta\right]^2\,.
\ee
Moreover, by defining the following superpotential
\be
W(\xi,\eta)=\frac{1}{\sqrt{2}}\,\left(\frac{\xi^2}{3}+\eta^2-1\right)\,\xi\,,
\ee
we are able to rewrite our potential $\widetilde{V}(\xi,\eta)$  as
\be \label{pot_2}
\widetilde{V}(\xi,\eta)=\frac{W_\xi^2}{2}+\frac{1}{2\,\beta^{\,2}}\,W_\eta^2\,,
\ee
where
\be \label{fod_01}
\beta=\sqrt{\frac{2}{\lambda-1}}\,, \qquad W_\xi=\frac{\partial W(\xi,\eta)}{\partial \xi}, \qquad W_\eta=\frac{\partial W(\xi,\eta)}{\partial \eta}.
\ee

Therefore, if $\lambda=3$ the analytical case studied by Trullinger and Subbaswamy in \cite{trull} is recovered naturally. 

Furthermore, the total energy for the fields configurations is such that
\be
E=\int_{-\infty}^{+\infty}\,d\rho\,\left[\frac{1}{2}\,(\xi_{\,\rho}^{\,2}+\eta_{\,\rho}^{\,2})+\frac{W_\xi^2}{2}+\frac{1}{2\,\beta^{\,2}}\,W_\eta^2\right]\,,
\ee
then, repeating the BPS procedure we find
\be
E=\int_{-\infty}^{+\infty}\,d\rho\,\left[\frac{1}{2}\,\left(\xi_\rho \mp W_{\xi}\right)^{\,2}+\frac{1}{2}\,\left(\eta_\rho \mp \frac{W_{\eta}}{\beta}\right)^{\,2} \pm \,\xi_{\rho}\,W_{\xi} \pm \,\eta_{\rho}\,\frac{W_{\eta}}{\beta}\right]\,.
\ee
So, if the first-order differential equations
\be \label{orb_a}
\xi_\rho=\pm\,W_\xi=\pm\,\frac{1}{\sqrt{2}}\,(\xi^2+\eta^2-1)\,,
\ee
\be \label{orb_b}
\eta_\rho=\pm\,\frac{W_\eta}{\beta}=\pm\,\sqrt{\lambda-1}\,\xi\,\eta\,,
\ee
are obeyed, we have the following effective energy
\be
E=\int_{-\infty}^{+\infty}\,d\rho\,\left(\xi_{\rho}\,W_{\xi}+\eta_{\rho}\,\frac{W_{\eta}}{\beta}\right)\,,
\ee
which can be rewritten as 
\be
E=E_{\,BPS}+\left(\frac{1}{\beta}-1\right)\,\int_{-\infty}^{+\infty}\,d\rho\eta_{\rho}\,W_{\eta}\,, 
\ee
where the BPS energy is simply
\be
E_{\,BPS}=\int_{-\infty}^{+\infty}\,d\rho\,\frac{d\,W}{d\,\rho}=W(\xi(\infty),\eta(\infty))-W(\xi(-\infty),\eta(-\infty))=\Delta\,W\,.
\ee

{\bf Thus, we can see that $E=E_{\,BPS}$ only if $\beta=1$. In order to find general configurations, let us compute the possible analytical solutions of the first-order differential equations \eqref{orb_a} and \eqref{orb_b}. One path to integrate such equations consists in rewrite them as
\be
\frac{d\,\xi}{d\,\eta}=\frac{1}{\left[2\,(\lambda-1)\right]^{1/2}}\,\frac{\xi^2+\eta^2-1}{\xi\,\eta}\,.
\ee
Now, using the new variable $\sigma=\xi^2-1$, the above equation takes the form
\be
\sigma_\eta=\sqrt{\frac{2}{\lambda-1}}\,\frac{\sigma+\eta^2}{\eta}\,.
\ee
Solving the previous differential equation, we conclude, after straightforward manipulations, that the relation between $\xi$ and $\eta$ is
\be \label{orb_d}
\xi^2=1+\frac{\beta}{2-\beta}\,\eta^2+c\,\eta^{\beta}\,,
\ee
where $c$ is an arbitrary integration constant. So, we directly see that the case $\beta=1$ (or $\lambda=3$), means
\be \label{orb_e}
\eta_\rho=\pm\,\eta\,(1+c\,\eta+\eta^2)^{1/2}\,,
\ee
and by integrating Eqs. \eqref{orb_a} and \eqref{orb_e} we determine the following solutions }
\be \label{sol_5}
\eta \left( \rho \right) =\frac{4}{e^{-\sqrt{2}\rho }-2c+e^{\sqrt{2%
}\rho}\left(c^{2} - 4\right) } \,,
\ee
\be \label{sol_6}
\xi \left( \rho \right) =\frac{-1+e^{2\, \sqrt{2}\rho}\left(c^{2}-4\right) }{1-2\,c\,e^{\sqrt{2}\rho}+e^{2\,
\sqrt{2}\rho}\left(c^{2}-4\right) }\,.
\ee

\begin{figure}[hb!]
\vspace{0.2cm}
\includegraphics[width=0.4\columnwidth]{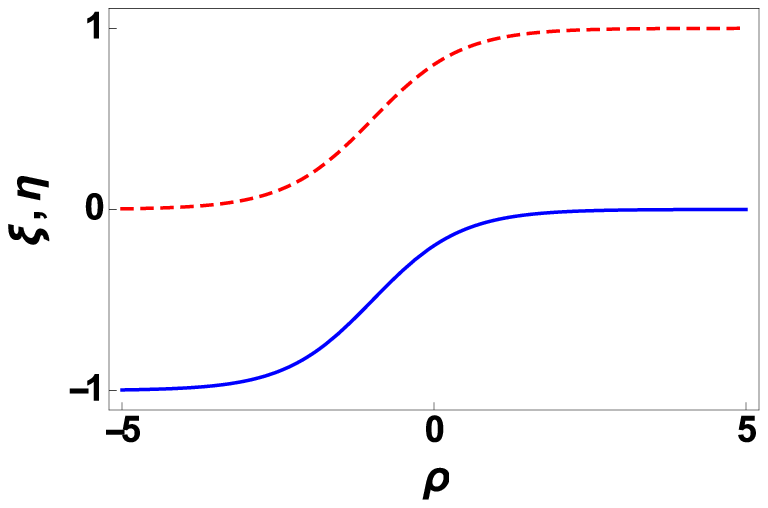}
\hspace{0.3cm}
\includegraphics[width=0.4\columnwidth]{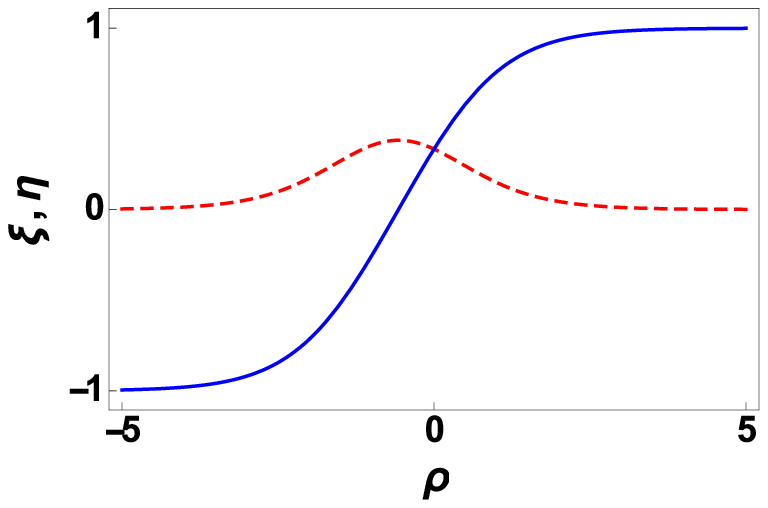}
\vspace{0.3cm}
\caption{In the left panel we present the critical solutions $\xi$ (solid blue curve) and $\eta$ (dotted red curve) for $\beta=1$ and $c=-2$. The right graphic shows the subcritical solutions $\xi$ (solid blue curve) and $\eta$ (dotted red curve) for $\beta=1$ and $c=-3$. }
\label{fig01}
\end{figure}

\begin{figure}[ht!]
\vspace{0.2cm}
\includegraphics[width=0.4\columnwidth]{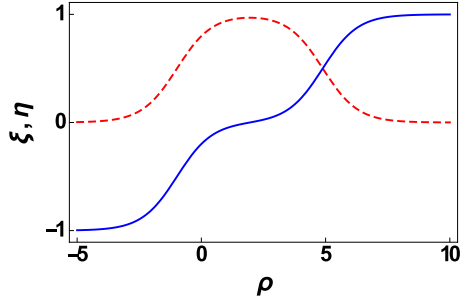}
\caption{This graphic shows the transition between critical and subcritical solutions. The solutions $\xi$ (solid blue double-kink) and $\eta$ (dotted red curve plateau-like lump) were plotted with $\beta=1$ and $c=-2.001$. }
\label{fig02}
\end{figure}

{\bf The above solutions with a general value of $c$ are new sets of configurations for the model proposed by Trullinger and  Subbaswamy in \cite{trull}, and they are graphically represented in Figs. \ref{fig01} and \ref{fig02}. In Fig. \ref{fig01} one can see two types of solutions: one called critical, where both graphics are kinks (left panel of Fig. \ref{fig01}), and one called subcritical where $\xi$ is a kink while $\eta$ is a lump (right panel of Fig. \ref{fig01}). Besides in Fig. \ref{fig02} we depicted the transition between the subcritical and the critical cases, where we have a double-kink defect for $\xi$ and a plateau-like lump for $\eta$.  Moreover, the solutions are divergent if $c>-2$. Defects like these were found for other two scalar fields models as one can see in \cite{alonso_02,dutra_plb}.}

Another set of analytical solutions can be determined if we take $\beta=4$ (or $\lambda=9/8$), leading to the orbit
\be \label{orbg}
\xi^2=1-2\,\eta^2+c\,\eta^4\,,
\ee
thus, from Eq. $(\ref{orb_b})$, we have 
\be \label{orb_f}
\eta_\rho=\frac{\sqrt{2}}{4}\,\eta\,(1-2\,\eta^2+c\,\eta^4)^{1/2}\,,
\ee
whose analytical solutions are
\be \label{sol_3}
\eta(\rho)=\,\frac{2\,e^{\frac{\rho }{2 \sqrt{2}}}}{\sqrt{1+4 e^{\frac{\rho }{\sqrt{2}}}-4 (c-1)\, e^{\sqrt{2} \rho}}}\,,
\ee
\be \label{sol_4}
\xi(\rho)=\frac{1+4\, (c-1)\, e^{\sqrt{2} \rho }}{1+4\, e^{\frac{\rho }{\sqrt{2}}}-4 \,(c-1)\, e^{\sqrt{2} \rho }}\,.
\ee
These two solutions clearly satisfy our first-order differential equations \eqref{orb_a} and \eqref{orb_f}, however, they are not solutions of the equations of motion investigated in \cite{trull}. {\bf Let us show this affirmative in more details considering the two-field Lagrangian density
$$
{\cal L}= -\frac{\xi_\rho}{2}-\frac{\eta_\rho}{2}-\frac{W_\xi^2}{2}-\frac{W_\eta^2}{2\,\beta^{\,2}}\,,
$$
whose equations of motion are}
\be \label{eqm_1}
\xi_{\rho\,\rho}-\left(W_\xi\,W_{\xi\,\xi}+\frac{W_\eta}{\beta^{\,2}}\,W_{\xi\,\eta}\right)=0\,,
\ee
\be \label{eqm_2}
\eta_{\rho\,\rho}-\left(W_\xi\,W_{\xi\,\eta}+\frac{W_\eta}{\beta^{\,2}}\,W_{\eta\,\eta}\right)=0\,.
\ee
As we saw before, the first-order differential equations for this model have the form
\be
\xi_\rho=W_\xi\qquad \eta_\rho=\frac{W_\eta}{\beta}\,,
\ee
so, by taking a derivative in respect to $\rho$ of the previous equations, we find
\be
\xi_{\rho\,\rho}=W_{\xi\,\xi}\,\xi_\rho+W_{\xi\,\eta}\,\eta_\rho\,;
\qquad
\eta_{\rho\,\rho}=\frac{1}{\beta}\left(W_{\eta\,\eta}\,\eta_\rho+W_{\eta\,\xi}\,\xi_{\rho}\right)\,,
\ee
which can be rewritten as
\be
\xi_{\rho\,\rho}-\left(W_{\xi\,\xi}\,W_\xi+\frac{W_{\eta\,\xi}}{\beta}\,W_{\eta}\right)=0\,;
\qquad
\eta_{\rho\,\rho}-\left(\frac{W_{\eta\,\xi}}{\beta}\,W_{\xi}+\frac{W_{\eta\,\eta}}{\beta^{\,2}}\,W_{\eta}\right)=0\,.
\ee
{\bf By comparing the previous results with Eqs. \eqref{eqm_1} and \eqref{eqm_2}, we observe that they are consistent only if $\beta=1$. Therefore, defects for $\beta \neq 1$ are not considered as physical solutions of $(\ref{in_eq1})$, once they do not satisfy the equations of motion of this model. So, we yield to the following issues:  What kind of models have the solutions presented in \eqref{sol_3}, and \eqref{sol_4}? If lumps and kinks exist in these new models, what are their physical importance? Furthermore, where can we find such models?}

\section{Transforming non-physical solutions in BPS states}
\label{sec_3}

In this section, we will answer the questions presented above. Then, we are going to construct effective two-field models with these analytical solutions and their correspondent orbit equation. The effective model is obtained via the application of an extension method developed in \cite{bls}. Such a method was successfully applied in the context of a cosmological model as one can see in \cite{ms_prd}, to build analytical three scalar field models \cite{sm_2018} and recently to derive new topological twiston-like defects for polyethylene molecules \cite{mbo_2018}.  In this approach, we are going to deal with the analytical solutions
\be \label{sol_1}
\eta(\rho)=\frac{2\,e^{\frac{\rho }{2\,\sqrt{2}}}}{\sqrt{1+4\,e^{\frac{\rho }{\sqrt{2}}}}} \,,
\ee
\be \label{sol_2}
\xi(\rho)=\frac{1}{\left(1+4\,e^{\frac{\rho }{\sqrt{2}}}\right)}\,,
\ee
where we choose $c=1$ in Eqs. \eqref{sol_3} and \eqref{sol_4} which are the critical solutions for $\beta=4$ (both solutions are kinks for this value of $c$). This critical behavior was chosen to prevent terms with rational exponents in our scalar potential. Therefore, the orbit Eq. \eqref{orbg} which relates fields $\eta$ and $\xi$ becomes
\be \label{def_1}
\eta=\sqrt{1-\xi}\,.
\ee
So, we are able to rewrite Eq. \eqref{orb_a} as
\be
\xi_\rho=\frac{\xi}{\sqrt{2}}\,(\xi-1)\,,
\ee
let us also note that by inverting Eq. \eqref{def_1}, we have
\be \label{func}
\xi=f(\eta)=1-\eta^2\,,
\ee
where we can call $f(\eta)$ as an orbit equation or as a deformation function, in analogy with the deformation method for one-field models introduced in \cite{blm}. With the previous ingredients Eq. \eqref{orb_f} is given by
\be
\eta_\rho=\frac{\eta}{\sqrt{8}}\,(1-\eta^2)\,.
\ee
 
{\bf Let us apply  the deformation function (and its inverse), to rewrite the first-order differential equations for $\xi(\rho)$, and $\eta(\rho)$ in three different but equivalent forms as follows}
\be
\xi_\rho=W_\xi(\xi)=W_\xi(\eta,\xi)=W_\xi(\eta), \qquad  \eta_\rho=W_\eta(\eta)=W_\eta(\eta,\xi)=W_\eta(\xi)\,\,.
\ee
Therefore, we can establish that an effective two scalar fields model obeys the following relation
\be \label{orb_c}
\xi_\eta=f_\eta=\frac{W_\xi(\xi\rightarrow\eta)}{W_\eta(\eta)}=\frac{a_1\,W_\xi(\eta)+a_2\,W_\xi(\xi ,\eta)+a_3\,W_\xi(\xi)+c_1\,g(\eta)+c_2\,g(\xi,\eta)+c_3\,g(\xi)}{b_1\,W_\eta(\eta)+b_2\,W_\eta(\xi ,\eta)+b_3\,W_\eta(\xi)}=\frac{\widetilde{W}_\xi(\xi,\eta)}{\widetilde{W}_\eta(\xi,\eta)}\,,
\ee
with the constraints
\be \label{vinc_1}
a_1+a_2+a_3=1\,,\qquad b_1+b_2+b_3=1 \qquad \text{ e } \qquad c_1+c_2+c_3=0\,,
\ee
and
\be \label{vinc_2}
b_2\,W_{\eta\,\xi}(\xi,\eta)+b_3\,W_{\eta\,\xi}(\xi)=a_1\,W_{\xi\,\eta}(\eta)+a_2\,W_{\xi\,\eta}(\xi,\eta)+c_1\,g_\eta(\eta)+c_2\,g_{\eta}(\xi,\eta)\,,
\ee
where the last constraint is a consequence of the property
\be
\widetilde{W}_{\xi\,\eta}(\xi,\eta)=\widetilde{W}_{\eta\,\xi}(\xi,\eta)\,.
\ee
So, the equations took as ingredients to construct the effective two-field models are 
\be
\xi_\rho=W_\xi(\xi)=\frac{\xi}{\sqrt{2}}\,(\xi-1) \qquad \xi_\rho=W_\xi(\xi,\eta)=\frac{1}{\sqrt{2}}\,\left(\xi^2+\eta^2-1\right)\qquad \xi_\rho=W_\xi(\eta)=-\frac{1}{\sqrt{2}}\,\eta^2\,\left(1-\eta^2\right)\,,
\ee

\begin{figure}[ht!]
\vspace{0.2cm}
\includegraphics[width=0.3\columnwidth]{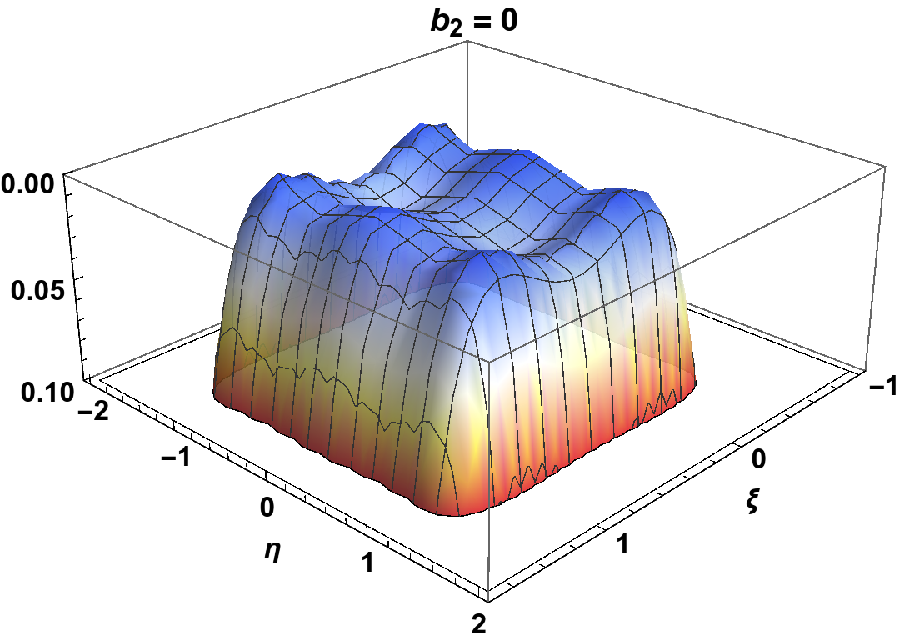}
\hspace{0.3cm}
\includegraphics[width=0.3\columnwidth]{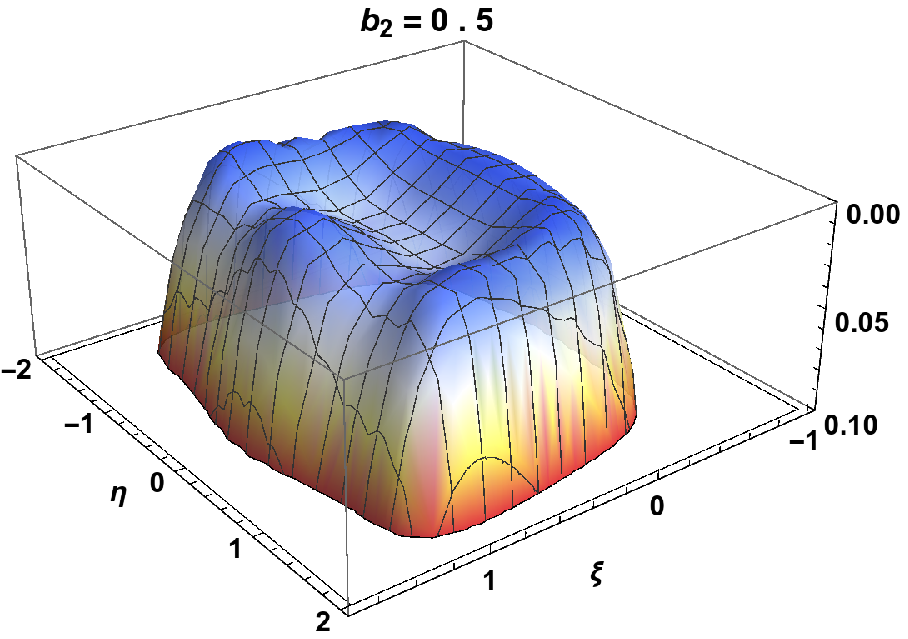}
\vspace{0.3cm}
\caption{These graphics unveil upside down views of effective potential $V(\xi,\eta)$, with $b_2=0$ (left panel), and $b_2=0.5$ (right panel).}
\label{fig1}
\end{figure}
\begin{figure}[ht!]
\vspace{0.2cm}
\includegraphics[width=0.3\columnwidth]{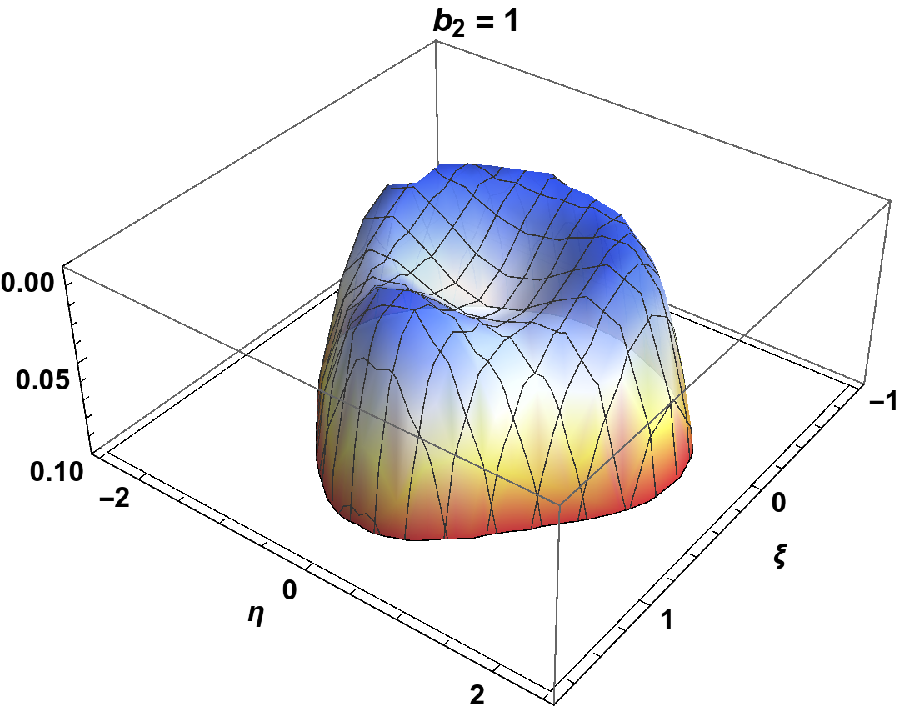}
\hspace{0.3cm}
\includegraphics[width=0.3\columnwidth]{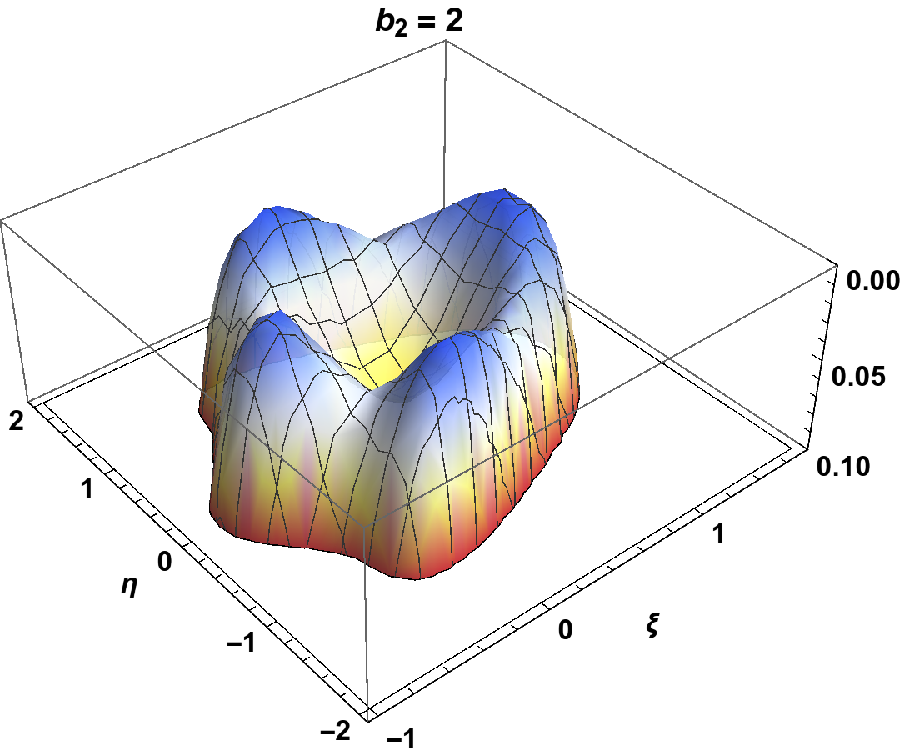}
\vspace{0.3cm}
\caption{These graphics unveil upside down views of  $V(\xi,\eta)$, with $b_2=1$ (left panel), and  $b_2=2$ (right panel).}
\label{fig2}
\end{figure}

\be
\eta_\rho=W_\eta(\eta)=\frac{1}{\sqrt{8}}\,\eta\,(1-\eta^2)\,,\qquad\eta_\rho=W_\eta(\eta,\xi)=\frac{1}{\sqrt{8}}\,\eta\,\xi\,,
\ee
where we choose $b_3=0$ to avoid terms with rational exponents in our scalar potential. Thus, by substituting such equations into Eq. \eqref{vinc_2}, we obtain 
\be
c_2\,g(\xi,\eta)=\frac{b_2}{4\,\sqrt{2}}\,\eta^2-\frac{a_2}{\sqrt{2}}\,\left(\xi^2+\eta^2-1\right)+\frac{a_1}{\sqrt{2}}\,\eta^2\,\left(1-\eta^2\right)\,,
\ee
where we worked with $c_1=0$. Now, using the deformation function Eq. \eqref{func}, we can rewrite the above equation as
\be
c_2\,g(\xi)=\frac{b_2}{4\,\sqrt{2}}\,(1-\xi)-\frac{a_2+a_1}{\sqrt{2}}\,(\xi^2-\xi)\,.
\ee
Then, by putting $g$ into Eq. \eqref{orb_c} and by integrating both $\widetilde{W}_\xi(\xi,\eta)$, and $\widetilde{W}_\eta(\xi,\eta)$ we find the effective superpotential 
\be \label{super01}
\widetilde{W}(\xi,\eta)=\frac{(1-b_2)}{2\,\sqrt{2}}\,\left(\frac{\eta^2}{2}-\frac{\eta^4}{4}\right)+\frac{b_2}{4\,\sqrt{2}}\,\eta^2\,\xi-\frac{1}{\sqrt{2}}\,\left(\frac{\xi^2}{2}-\frac{\xi^3}{3}\right)-\frac{b_2}{4\,\sqrt{2}}\left(\xi-\frac{\xi^2}{2}\right)\,,
\ee
which can be used to tailor different potentials $V(\xi,\eta)$ with the form
\be \label{pot01}
V(\xi,\eta)=\frac{\widetilde{W}_\xi^{\,2}}{2}+\frac{\widetilde{W}_\eta^{\,2}}{2}\,.
\ee
The behavior of these potentials are shown in Figs.  $\ref{fig1}$ and $\ref{fig2}$. There we observe that constant $b_2$, which came from the extension procedure, allows the construction of different sets of minima for our effective potential. Besides, the mentioned constant also is responsible to deform the potential as we clearly see in Fig $\ref{fig2}$. A simple example of new model consists in the case $b_2=1$, where the potential is such that
\be
V_1(\xi,\eta)=\frac{1}{64} \left[4\, \eta^2 \,\xi^2+\left(\eta^2-(1-\xi ) (1+4\,\xi )\right)^2\right]\,.
\ee
In this specific case, the correspondent equations of motion are written as
\be
\xi_{\rho\,\rho}=\frac{1}{32}\, \left[3+\xi -3 \eta ^2 (1-4\,\xi )-4 \xi ^2 (9-8\,\xi )\right]\,,
\ee
\be
\eta_{\rho\,\rho}=\frac{1}{16} \,\eta \, \left(\eta ^2-3 \xi +6 \xi ^2-1\right)\,.
\ee
At this point, it is important to remark that the fields configurations given by Eqs. \eqref{sol_1}, and \eqref{sol_2} satisfy the above equations. In addition, we can note that this pair of coupled equations is different from  Eqs. \eqref{sordem_1} and \eqref{sordem_2}. The BPS energy for the different models is computed considering
\be
E_{BPS}=\widetilde{W}(+\infty,+\infty)-\widetilde{W}(-\infty,-\infty)=\frac{7}{24 \sqrt{2}}\,,
\ee
{\bf unveiling that the kink-like solutions form a topological sector which is stable, besides, all models are degenerated in respect to this sector. Therefore, in this section, we have shown that it is possible to obtain a new class of BPS models using a set of physical solutions from \cite{trull}.}

\section{Application in cosmology}
\label{sec_3_1}

An interesting application of the new analytical models found in the last section is in the context of hybrid inflation. There, the standard Einstein-Hilbert Lagrangian is coupled with a two real scalar fields Lagrangian density. This procedure is adopted in order to describe an Universe passing through different inflationary eras and dominated by dark energy for later values of time. 
Let us implement such a formalism using the action
\bn \label{sec5_eq1}
&& \nonumber
S=\int\,d^{\,4}\,x\,\sqrt{-g}\,\left(-\frac{R}{4}+{\cal L}(\xi,\partial_{\,\mu}\xi,\eta,\partial_{\,\mu}\eta)\right)\,; \\ 
&&
{\cal L}=\frac{1}{2}\,\partial_{\,\mu}\,\xi\,\partial^{\,\mu}\,\xi+\frac{1}{2}\,\partial_{\,\mu}\,\eta\,\partial^{\,\mu}\,\eta-V(\xi,\eta)\,,
\en
with $\xi=\xi(t)$, $\eta=\eta(t)$, $4\,\pi\,G=1$, $c=1$, and metric signature $(+,-,-,-)$. Once we are dealing with fields which depend only on time variable, we are going to take
\be
\rho \rightarrow \, t+t_0\,,
\ee
in the expressions found in the last section.

By minimizing the action $(\ref{sec5_eq1})$ in respect to the metric we have
\be \label{sec5_eq2}
R_{\,\mu\,\nu}-\frac{1}{2}\,g_{\,\mu\,\nu}\,R=2\,T_{\,\mu\,\nu}\,,
\ee
where $T_{\,\mu\,\nu}$ is denominated as energy-momentum tensor and it is such that
\be \label{sec5_eq3}
T_{\,\mu\,\nu}=2\,\frac{\partial\,{\cal L}}{\partial\,g^{\,\mu\,\nu}}-g_{\,\mu\,\nu}\,{\cal L}\,.
\ee
The energy-momentum tensor has  $(\rho_s,-p_s,-p_s,-p_s)$ as its components, with $\rho_s$, and $p_s$ as the density and the pressure related with the scalar fields. 

The last ingredients enable us to find that
\be \label{sec5_eq4}
\rho_s=\frac{\dot{\xi}^{\,2}}{2}+\frac{\dot{\eta}^{\,2}}{2}+V\,; \qquad
p_s=\frac{\dot{\xi}^{\,2}}{2}+\frac{\dot{\eta}^{\,2}}{2}-V\,. \\ 
\ee
Besides, if we choose to work with flat Friedmann-Robertson-Walker metric, we yield to the Friedmann equations
\be \label{sec5_eq5}
H^{\,2}=\frac{2}{3}\,\rho_s\,; \qquad 
\dot{H}+H^{\,2} =-\frac{1}{3}\,\left(\rho_s+3\,p_s\right)\,,; \qquad H=\frac{\dot{a}}{a}\,,
\ee
where $H$ is called Hubble parameter. The last results can be rewritten in a more convenient way as follows 
\be \label{sec5_eq6}
H^{\,2}=\frac{1}{3}\,\left(\dot{\xi}^{\,2}+\dot{\eta}^{\,2}+2\,V\right)\,; \qquad
\dot{H}=-\left(\dot{\xi}^{\,2}+\dot{\eta}^{\,2}\right)\,.
\ee

Another parameter which is important for cosmological phenomenology is the Equation of State (EoS) parameter, whose explicit form is
\be \label{sec5_eq6_1}
\omega=\frac{p_s}{\rho_s}=\frac{\dot{\xi}^{\,2}+\dot{\eta}^{\,2}-2\,V}{\dot{\xi}^{\,2}+\dot{\eta}^{\,2}+2\,V}\,.
\ee
It is relevant to point that such a parameter is measured by collaborations like PLANCK and Dark Energy Survey \cite{planck18,des_18}, consequently, the EoS parameter consists in an excellent test to verify the validity of a given model.  

In order to derive analytical cosmological models, we use the first-order formalism, which is based on the constraint 
\be \label{sec5_eq7}
H=-W(\xi, \eta)\,,
\ee
yielding to
\be \label{sec5_eq8}
\dot{H}=-W_{\,\xi}\,\dot{\xi}-W_{\,\eta}\,\dot{\eta}\,.
\ee

Thus, by taking $H$, and $\dot{H}$ into (\ref{sec5_eq6}) we find the first-order differential equations
\be \label{sec5_eq9}
\dot{\xi}=W_{\,\xi}\,; \qquad \dot{\eta}=W_{\,\eta}\,,
\ee 
as well as the cosmic potential
\be \label{sec5_eq10}
V=\frac{3}{2}\,W^{\,2}-\frac{1}{2}\,\left(W_{\,\xi}^{\,2}+W_{\,\eta}^{\,2}\right)\,.
\ee 

By minimizing the action (\ref{sec5_eq1}) in respect to the fields, we derive the equations of motion
\be \label{sec5_eq11}
\ddot{\xi}+3\,H\,\dot{\xi}+V_{\xi}=0\,; \qquad \ddot{\eta}+3\,H\,\dot{\eta}+V_{\eta}=0\,,
\ee
which need to be satisfied by the solutions of the first-order equations presented in (\ref{sec5_eq9}).

\begin{figure}[h!]
\centering
\includegraphics[width=0.3\columnwidth]{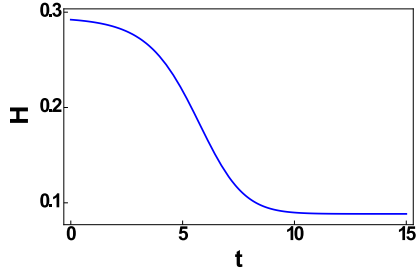}
\caption{Time evolution of the analytical Hubble parameter derived from our two-field model. There is a small step close to $t=0$ indicating a primordial expansion era which smoothly evolves to another expansion era for later values of time. }
\label{fig3}
\end{figure}

After these generalities, we are ready to apply our model in such a cosmological scenario. The analytical solutions which are going to satisfy the first-order equations (\ref{sec5_eq9})  are
\be \label{sec5_sol_1}
\xi(t)=\frac{1}{\left(1+4\,e^{\frac{t+t_0 }{\sqrt{2}}}\right)}\,; \qquad \eta(t)=\frac{2\,e^{\frac{t+t_0 }{2\,\sqrt{2}}}}{\sqrt{1+4\,e^{\frac{t+t_0 }{\sqrt{2}}}}} \,,
\ee
then, by taking these expressions together with the superpotential presented in (\ref{super01}), we obtain
\be \label{sec5_eq13}
H(t)=\frac{1}{24 \sqrt{2}}\left(3 \,b_2+\frac{15}{\left(4 e^{\frac{t+t_0}{\sqrt{2}}}+1\right)^2}-\frac{8}{\left(4 e^{\frac{t+t_0}{\sqrt{2}}}+1\right)^3}-3\right)\,,
\ee
as the Hubble parameter. The behavior of $H$ is shown in Fig. \ref{fig3}, where we worked with $b_2=2$, and $t_0=-8$. 

\begin{figure}[h!]
\centering
\includegraphics[width=0.3\columnwidth]{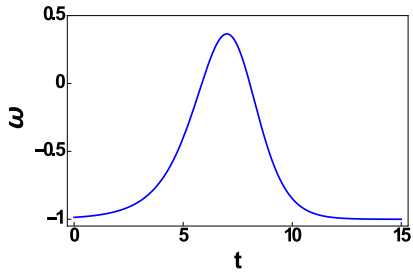}
\caption{Time evolution of the analytical EoS parameter derived from our two-field model. The picture was depicted with $b_2=2$ and $t_0=-8$. There we observe two expansion eras where $\omega \approx -1$ and that the parameter has maximum close to $\omega \approx 1/3$.}
\label{fig4}
\end{figure}

With the Hubble parameter in hands we are able to determine $V$  (\ref{sec5_eq10}), and the EoS parameter (\ref{sec5_eq6_1}) as
\begin{widetext}
\ben \label{sec5_eq14}
&& 
V=\frac{1}{768} \bigg[\left(6 \,b_2\, \left(\eta ^2-1\right) \xi +3\, (b_2-1)\, \eta ^2 \, \left(\eta ^2-2\right)+3\, (b_2-4)\, \xi ^2+8 \,\xi ^3\right)^2 \\ \nonumber
&&
-48 \, \left((b_2-1) \,  \left(\eta ^2-1\right)\,\eta +b_2 \, \eta \, \xi \right)^2-12\, \left(b_2 \,\left(\eta ^2+\xi -1\right)+4\, (\xi -1)\, \xi \right)^2\bigg]\,,
\een
\ben \label{sec5_eq15}
&&
\omega=\bigg[55296\, (b_2-1)^2\, e^{\frac{5 (t+t_0)}{\sqrt{2}}}-48\, \left(45 \,b_2^2+42\, b_2-236\right)\, e^{\sqrt{2} (t+t_0)}+36864 \,(b_2-1)^2\, e^{3 \sqrt{2} (t+t_0)} \\ \nonumber
&&
+\,24\, (9\, b_2\, (b_2+2)-8)\, e^{\frac{t+t_0}{\sqrt{2}}}+768\, (b_2 \,(15 \,b_2-4)-99) e^{\frac{3 (t+t_0)}{\sqrt{2}}}+3840\, (3\, b_2 (3\, b_2-4)-29) e^{2 \sqrt{2} (t+t_0)}+(3\, b_2+4)^2\bigg] \\ \nonumber
&&
\times\,\bigg[\left(192\, (b_2-1) e^{\frac{3 (t+t_0)}{\sqrt{2}}}+144\, (b_2-1) e^{\sqrt{2} (t+t_0)}+12\, (3 b_2+2) e^{\frac{t+t_0}{\sqrt{2}}}+3\, b_2+4\right)^2\bigg]^{\,-1}
\een
\end{widetext}
The features of the EoS parameter can be visualized in Fig. \ref{fig4}. Moreover, we can take $V$ (\ref{sec5_eq14}) together with the superpotential $W$ (\ref{super01}), and the solutions (\ref{sec5_sol_1}) to verify that the equations of motion (\ref{sec5_eq11}) are indeed satisfied.

\section{Cosmological interpretations}
\label{sec_3_2}
In this section, we study in details the behavior of the cosmological parameters found with the new class of models determined in section \ref{sec_3}. The inflationary model establishes that the Hubble parameter is approximately constant during the initial phase of our Universe \cite{ryden,dodelson}, and this feature can be clearly observed in Fig. \ref{fig3} for $t\approx 0$. In this same era, the EoS parameter is such that $\omega < -1/3$, which is shown in Fig. \ref{fig4}.

  After the first inflationary era, $H$ should decrease with time, and during its evolution, the EoS parameter enters into the radiation era, where the density of the Universe is three times larger than its pressure, which means $\omega = 1/3$ \cite{ryden,dodelson}. Finally, after the radiation era, the Universe passes through a second expansion era where $H \approx cte$. In this expansion era, the EoS parameter is expected to be $\omega \approx -1$, as established by experiments such as PLANCK and Dark Energy Survey \cite{planck18,des_18}. The mentioned value of $\omega$ characterizes the dark energy age. As we realize Figs. \ref{fig3} and \ref{fig4} present all these desired features. 

  It is really interesting that our analytical model is able to describe all the different eras expected from the inflationary theory, besides, we also point that these special behaviors are related with the value of the $b_2$ constant. If this constant is too different of $b_2=2$, then the features of the cosmological parameters are no longer compatible with the description of our Universe. As we realize in Figs. \ref{fig1} and \ref{fig2}, this constant is responsible to deform the scalar potential, so, we can establish a direct connection between this deformation constant and the physical behavior of the cosmological parameters.

\section{Final Remarks}
\label{sec_4}

In this paper, we studied the model proposed by Trullinger and Subbaswamy in the BPS perspective. We were able to generalize the class of solutions introduced in \cite{trull}, presenting the double kink and the plateau-like lump. Furthermore, we determined non-physical defects related with $\beta=4$. {\bf The defects characterized as non-physical do not satisfy the equations of motion of a given system, unlike the BPS ones.  In order to find a theory where these solutions are physically accepted, we applied the extension method to construct new two-field BPS models.    
Then, we tailored a procedure able to connect solutions which came from a non-standard BPS potential with new sets of BPS models. Such an approach shows a new connection between scalar fields models, whose bridge is the first-order differential equations for non-standard BPS models. }

 {\bf Moreover, the models derived in the last section were built with two kink-like solutions whose asymptotic behaviors correspond to the vacua values of the potential $V$, presented in $(\ref{pot01})$. So, it means that the kink-like solutions used in this approach are domain walls of such models.  The method can be repeated combining a kink with a lump defect, resulting in analytical models composed by domain walls with internal structure.} We believe that such methodology extends the studies concerning the BPS method and can be applied to other two-field models presented in the literature, such as the Montonen one introduced in \cite{montonen}. 

As a matter of applicability of our methodology, we used the results obtained in section \ref{sec_3} in the context of hybrid cosmological models. There we have succeeded in deriving analytical cosmological parameters which describe the observed Universe. It is remarkable that our model presented two different expansion eras and also an EoS parameter which is compatible with the most recent data sets from PLANCK and from Dark Energy Survey collaborations \cite{planck18,des_18}.

Furthermore, it is important to highlight that the approach applied in the present work can be very powerful to investigate different subjects, such as the generation of coherent structures after cosmic inflation \cite{PRD-83-2011}, the dynamics of oscillons configurations \cite{PRD2015, AHEP, JHEP2018}, braneworld theories with internal structure \cite{PLB97, PRD2009, CQG2011}, the nonlinear sigma model \cite{PRL2008}, Lorentz and symmetry breaking systems \cite{EPJC2014, AOP2016}, and alternative theories of gravity \cite{EPJC2016, EPJC2016-R}.

\acknowledgments
The authors would like to thank Capes and CNPq (Brazilian agencies) for financial support. RACC is partially supported by FAPESP (Foundation for Support to Research of the State of S\~ao Paulo) under grants numbers 2016/03276-5 and 2017/26646-5. The authors also would like to thank the anonymous referees for their comments and suggestions which enhanced the quality of this work.

\end{document}